\documentclass[10pt,letterpaper,twocolumn]{revtex4-1} %% two column, final layout

\usepackage[draft]{hyperref}
\usepackage{graphicx}

\begin{document}

\title{Generation of high-order optical vortices using directly machined spiral phase mirrors}

\author{Geoff Campbell, Boris Hage, Ben Buchler and Ping Koy Lam$^{*}$}

\address{
Centre for Quantum Computation and Communication Technology \\ Department of Quantum Sciences, The Australian National University, Canberra, Australia
\\
$^*$Corresponding author: ping.lam@anu.edu.au
}

\begin{abstract}We report on the generation of high-order optical vortices by spiral phase mirrors. The phase mirrors are produced by direct machining with a diamond tool and are shown to produce high-quality optical vortices with topological charges ranging from 1 to 1000 at a wavelength of 532 nm. The direct machining technique is flexible and offers the promise of high-precision, large-diameter spiral phase mirrors that are compatible with high optical powers.\end{abstract}

\maketitle

\section{Introduction}

Optical vortices are light beams that carry orbital angular momentum and are characterised by a phase that increases azimuthally about a singularity at the centre of the beam\cite{Coullet:1989p2463}. These beams can be generated directly in a cylindrically symmetric laser cavity. It is, however, simpler in practice to approximate them by illuminating diffractive optics with more readily available laser modes\cite{Heckenberg:1992p4137,Beijersbergen:1994p2791,Kennedy:2002p3945}.

A variety of possible applications for optical vortices\cite{FrankeArnold:2008p3716} have driven research into their production. These applications include optical trapping and manipulation of colloidal particles\cite{Grier:2003p844} and cold atoms\cite{Kuga:1997p5064}, the generation of Bose-Einstein condensate vortices\cite{Natarajan:2006p4907}, phase-contrast microscopy\cite{Bernet:2006p3400}, image filtering for extra-solar planet detection\cite{Lee:2006p3375}, stimulated emission depletion (STED) microscopy\cite{Willig:2006p3766} and the encoding of optical quantum information\cite{Mair:2001p2490}.

The considerable utility of optical vortices has driven the development of several methods for their production. Most commonly, a phase shift of the form $e^{i l \phi}$ is applied to an incident laser beam where $l$ is the topological charge and $\phi$ is the azimuthal coordinate about the beam centre in the transverse plane.

One method to apply the required phase shift is to use a computer-controlled spatial-light modulator (SLM)\cite{Sundbeck:2005p2535}. This approach has the considerable advantages of being flexible and allowing the phase profile to be dynamically varied on 10 ms timescales. SLM's, however, suffer from high cost and limited spatial and phase resolution. Conversion efficiencies can be quite high for low-order optical vortices but begin to suffer from resolution limits as the order is increased.

Another approach is to use lithographic processes to construct a phase mask\cite{Kennedy:2002p3945,Moh:2006p2223,Kotlyar:2008p4053}. These phase masks have restrictions in terms of phase resolution, spatial resolution, maximum phase shift and useable area, although advanced lithographic techniques can relax many of these restrictions.

Here, we report on the production by direct machining of spiral phase mirrors (SPMs) that are suitable for generating optical vortices. We demonstrate that the resultant phase mirrors can deterministically produce high-quality optical vortices over a large range of topological charges from 1 to 1000. These mirrors can be produced economically and the size of the optics is restricted only by the geometry of the machining enclosure.

\section{Fabrication}

The SPMs are machined from aluminium using a single-crystal diamond tool. The machining is performed by an ultra-precision lathe, produced by Moore Nanotechnology. The tool position relative to the part surface is adjusted while the part is rotated at a constant rate on the spindle. High-resolution position encoders ensure that the tool is positioned with nanometric precision for each point on the part surface.

\begin{figure}
\centerline{\includegraphics[width=\columnwidth]{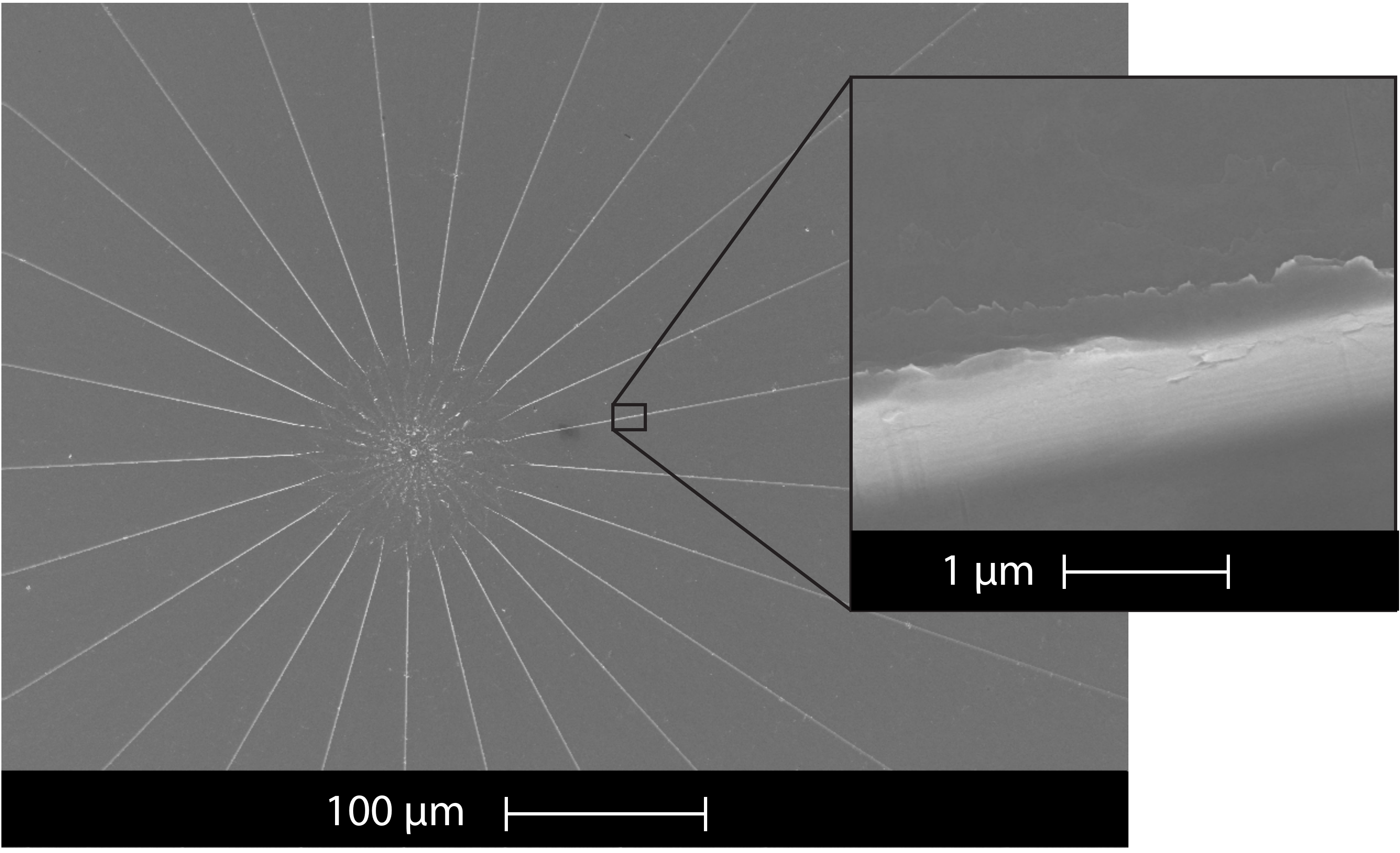}}
\caption{Scanning electron microscope image of the central region of a charge 100 spiral phase mirror. The inset shows the edge of one of the steps.}
\label{SEM}
\end{figure}

The cutting speed is limited by the acceleration of the lathe's linear axes, although we can relax this restriction by breaking the SPM into segments. Each segment is a linear ramp with a total height corresponding to a $2n\pi$ phase shift. Figure \ref{SEM} shows an SEM image of a charge 100 SPM composed of 25 segments, each of which imparts an $8\pi$ phase ramp. For the charge 100 SPM, we estimate that the central defect accounts for 0.006\% of the total mirror surface, while the sharp jumps between segments collectively account for approximately 3\% of the total area.

For our analysis, we produced 25 mm diameter optics, which required cutting times ranging from 30 minutes to two hours for charges of 1 to 100, respectively. The size of the optic can be scaled without losing quality or increasing cutting time simply by selecting a diamond tool of an appropriate radius.

\section{Results}

\begin{figure}
\centerline{\includegraphics[width=\columnwidth]{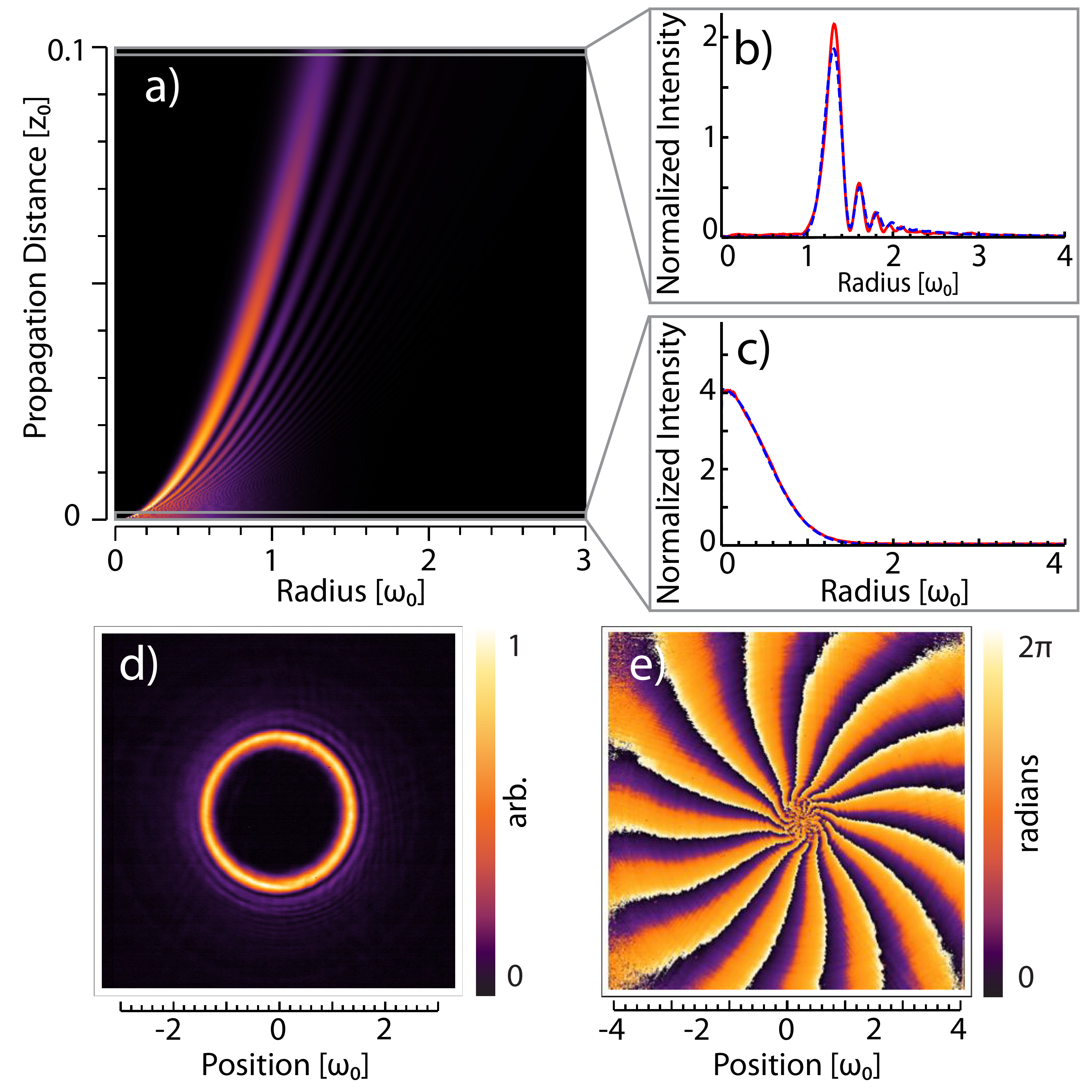}}
\caption{a) Calculated intensity of the mode produced by illuminating a charge 15 spiral phase mirror with a collimated TEM$_{00}$. b), c) Measured (solid red trace) and calculated (dashed blue trace) radial intensity profiles after propagation and at the location of the SPM, respectively. d) Measured intensity profile for a charge 15 optical vortex imaged in the far-field. e) Measured phase of the same vortex imaged in the near-field.}
\label{propagation}
\end{figure}

To verify the performance of our phase plate, we made transverse beam profile measurements that can be compared to theoretical models. The illumination of an SPM with a TEM$_{00}$ mode results in a beam profile that can be expressed as a hypergeometric Gaussian mode (HyGG)\cite{Karimi:2007p3929}. These modes have a cylindrically symmetric amplitude and propagate from a Gaussian profile at the SPM into a ring-shaped profile in the far-field. Figure \ref{propagation} a) shows the calculated radial amplitude of a propagating charge 15 HyGG mode immediately after it is created.

Experimental intensity profiles are obtained by illuminating machined SPMs with a 532 nm laser beam that has been transmitted through a mode-cleaning cavity locked to the TEM$_{00}$ mode. The resulting intensity patterns are recorded on a CCD camera. The radial intensity distribution is then obtained by fitting an interpolating function to the raw pixel data and integrating the resultant function azimuthally in polar co-ordinates. The experimental data is normalised such that the area under the curve is equal to that of the normalised HyGG integrated over the same region.

Figures \ref{propagation} b) and c) show the calculated (dashed blue) and measured (solid red) radial intensity profiles after propagation through 0.096 Rayleigh lengths and at the location of the SPM, respectively. The beam waist of the input Gaussian mode is 450 $\mu m$ and the propagation distance has been corrected from a measured 9.5 cm to 11.5 cm, a discrepancy which we attribute to the SPM not being located exactly at the beam waist. The excellent agreement of the model with our data indicates that the SPM is performing as designed. Figure \ref{propagation} d) shows the intensity of a charge 15 optical vortex that has been imaged in the far-field. The phase of the vortex in the near-field, figure \ref{propagation} e), has been obtained by interfering the vortex with a collimated Gaussian beam and scanning the relative phase of the two beams. CCD images of the interference patterns are captured and the phase profile is obtained by fitting a sine function to the intensity recorded at each pixel.

\begin{figure}
\centerline{\includegraphics[width=\columnwidth]{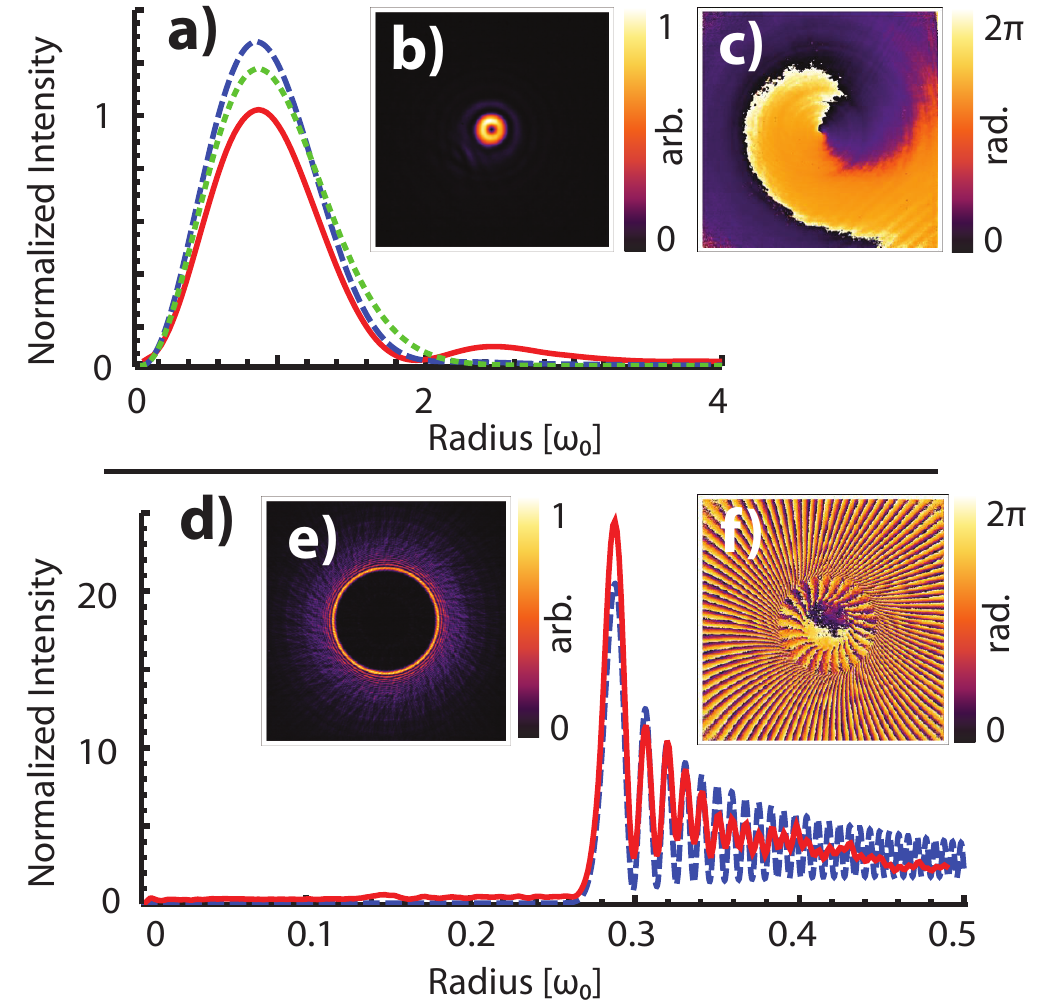}}
\caption{a) Measured (solid red) and calculated (long-dashed blue) radial intensity profiles for a charge 1 optical vortex. b) . For comparison, a charge 1 Laguerre-Gaussian mode is also shown (short-dashed green). c) The phase of the charge 1 vortex, imaged in the near-field. d) Measured (solid red) and calculated (dashed blue) intensity profiles for a charge 100 optical vortex, which is shown in e). f) Measured phase of the vortex in the same plane as the intensity measurement.}
\label{charge1and100}
\end{figure}

We verify that the machined SPMs produce high-quality vortices over a range of topological charges by examining the intensity and phase profiles of charge 1 and charge 100 vortices. Figure \ref{charge1and100} (a-c) shows the results for the charge 1 vortex. Due to the slow evolution of the charge 1 vortex, imaging optics were used to record the intensity pattern and the propagation distance which was used for the calculated intensity was selected to fit the measured profile.

Figure \ref{charge1and100} (d-f) shows the results for a charge 100 vortex. There is some deviation from the calculated intensity in the peripheral fringes; however, the primary feature is in good agreement with the calculated intensity distribution. As with the charge 15 analysis, the propagation distance was corrected to account for the SPM not being located at the beam waist. The phase profile shows three distinct regions. The central region has zero charge and originates from the undiffracted portion of the beam. An intermediate region shows a charge of 25, which we believe is an artefact of the 25 step design used to create the SPM. The inner region is, however, of little consequence because the outer region, beyond a radius of 0.25 $\omega_0$, contains 98\% of the optical power and shows the intended charge 100 optical vortex.

\begin{figure}
\vspace{6 mm}
\centerline{\includegraphics[width=\columnwidth]{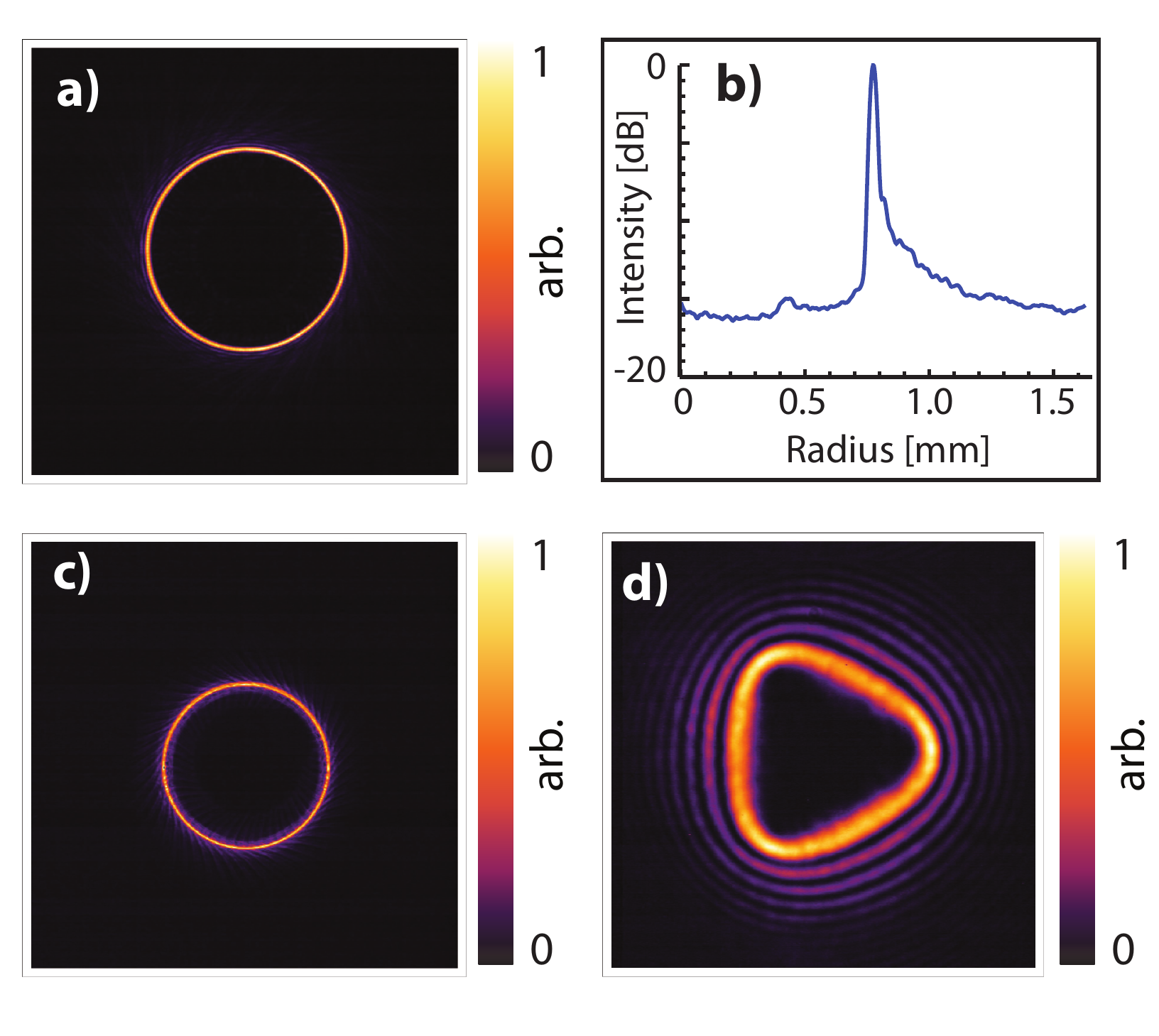}}
\caption{a) Intensity profile of the charge 100 optical vortex, imaged in the plane featuring the narrowest ring width. b) The same profile, integrated over the azimuthal coordinate and shown on a Log$_{10}$ scale. c) Intensity profile of a charge 1000 optical vortex. d) The vortex produced by a charge 15 SPM with a $2/3\pi$ period modulation of the phase ramp.}
\label{highcharge}
\end{figure}

The intensity distribution of the same charge 100 vortex, imaged in a plane that features the smallest observed ring thickness, is shown in figures \ref{highcharge} a) and b). The minimal undiffracted order, even at high charges, is notable when compared to SLM-produced vortices\cite{Sundbeck:2005p2535}.

In an attempt to find the limits of the direct machining process, we manufactured a charge 1000 SPM; however, our ability to properly characterise the vortices degraded after a charge of roughly 100 due to limited CCD resolution. Figure \ref{highcharge} c) shows the measured intensity profile of the charge 1000 optical vortex. 

Spiral phase mirrors that can be manufactured by this technique are not limited to linear azimuthal phase ramps. Modulated phase profiles\cite{Curtis:2003p2447} can be produced provided that the modulation is slow compared to the loop bandwidth of the positional feedback on the lathe. To demonstrate this flexibility, we cut a charge 15 optical vortex with a sinusoindally modulated phase ramp, shown in figure \ref{highcharge} d). The modulation has a period of $2/3 \pi$ and a depth of 0.1. It is also possible to cut the SPMs on a curved surface. In this way, SPMs could be created on spherical or even aspherical mirrors to provide integrated focusing optics.

\section{Conclusions}

We have demonstrated that the direct machining of spiral phase mirrors using a diamond tool can produce high-quality, high-charge optical vortices. In addition to low cost, technical simplicity and flexibility, this fabrication method has the advantage of being applicable to a variety of metallic and soft crystalline optical materials, which may allow the production of transmissive as well as reflective vortex producing optics.

\section*{Acknowledgments}

We thank A. Luiten, A. G. White, M. Hosseini and N. Devlin for fruitful discussion. We gratefully acknowledge the technical support provided by T. Stewart and E. Alarie from Moore Nanotechnology. This research is supported by the Australian Research Council (project numbers CE110001027 and DP1092891).

%\pagebreak

\end{document}